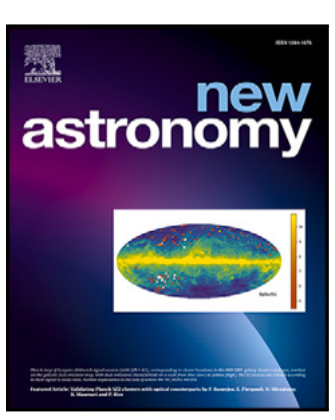

# Precision near-IR spectroscopy for understanding AGN physics and shed light on the $H_0$ tension ☆

Matilde Signorini [a,b,*], Vardha N. Bennert [c], Elena Dalla Bontà [e,f,g], Federica Ricci [d], Tommaso Treu [h], Lizvette Villafaña [c]

[a] ESA European Space Research and Technology Centre (ESTEC), Keplerlaan 1, Noordwijk, 2201 AZ, The Netherlands
[b] INAF - Osservatorio Astrofisico di Arcetri, Largo Enrico Fermi 5, Florence, I-50125, Italy
[c] Physics Department, California Polytechnic State University, 93407, San Luis Obispo CA, USA
[d] Dipartimento di Matematica e Fisica, Universit'a degli Studi di Roma Tre, via della Vasca Navale, 84, 00146, Roma, Italy
[e] Dipartimento di Fisica e Astronomia "G. Galilei", Università di Padova, Vicolo dell'Osservatorio 3, I-35122, Padova, Italy
[f] INAF – Osservatorio Astronomico di Padova, Vicolo dell'Osservatorio 5, I-35122, Padova, Italy
[g] Jeremiah Horrocks Institute, University of Central Lancashire, PR1 2HE, Preston, UK
[h] University of California, Department of Physics and Astronomy, 430 Portola Plaza, 90095, Los Angeles, CA, USA

ABSTRACT

The persistent tension between early- and late-Universe measurements of the Hubble constant ($H_0$) remains one of the most significant challenges in modern cosmology. The Spectroastrometry and Reverberation Mapping (SARM) method offers a promising, calibration-independent approach to address this issue by combining time-delay measurements of the Broad-Line Region (BLR) with interferometric angular size determinations. Current implementations of SARM, however, are limited by the difficulty of performing near-infrared reverberation mapping (RM) on the same emission lines observed by GRAVITY, restricting applications to only a few bright AGN. We propose using the capabilities of SHARP, the next-generation near-infrared spectrograph for the Extremely Large Telescope (ELT), to overcome these limitations. SHARP's sensitivity and multi-object spectroscopy will enable (1) efficient long-term monitoring of existing GRAVITY targets with minimal time investment, and (2) systematic RM campaigns for the fainter AGN that will be observed by GRAVITY+. These advances will give us precise infrared lags for tens of AGN, enabling geometric distance measurements and a robust, calibration-free determination of $H_0$. Beyond cosmology, SHARP will allow detailed studies of BLR structure and kinematics in the infrared, advancing our understanding of AGN physics and with repercussions on the measurements of Supermassive Black Holes (SMBH) masses.

## 1. Introduction

Active Galactic Nuclei (AGN) are powered by accretion of gas onto supermassive black holes (SMBHs) with masses in the range $10^6$–$10^{10}$ $M_\odot$, residing at the centre of galaxies, and with a luminosity that can outshine the stellar emission of their host galaxies across much of the electromagnetic spectrum. AGN are thought to co-evolve with their host galaxy, as shown by the observational evidence of the scaling relation between SMBH masses and host galaxies properties (e.g. Magorrian et al., 1998; Gebhardt et al., 2000). In unobscured (type 1) AGN, a defining observational feature is the presence of broad permitted emission lines with widths of several thousand km s$^{-1}$. These lines originate in the Broad-Line Region (BLR), a compact (few light-days to light-months) region of dense gas gravitationally bound to the SMBH and photoionised by the accretion disc. The BLR is spatially unresolved for most AGN, as its characteristic size corresponds to light-days to light-months, far below the angular resolution of direct imaging even with the largest telescopes. Reverberation Mapping (RM) (Blandford and McKee, 1982; Wandel et al., 1999; Vestergaard, 2002) provides a powerful indirect method to probe these regions. RM exploits the intrinsic variability of AGN: by monitoring the time-delayed response of the BLR emission-line flux to variations in the ionizing continuum, RM yields a measurement of the characteristic BLR radius, $R_{\rm BLR} \simeq c\,\tau$, with $c$ the speed of light and $\tau$ the time lag between the continuum and BLR response. Combined with the velocity width $\Delta V$ of the broad emission line, this enables an estimate of the black hole mass through

☆ This article is part of a Special issue entitled: 'SHARP Science Book' published in New Astronomy.
* Corresponding author at: ESA European Space Research and Technology Centre (ESTEC), Keplerlaan 1, Noordwijk, 2201 AZ, The Netherlands.
*E-mail address:* matilde.signorini@esa.int (M. Signorini).

the virial relation

$$M_{\rm BH} = f\,\frac{R_{\rm BLR}\,\Delta V^2}{G}, \tag{1}$$

where $f$ is the virial factor, which comes from the combined effect of the geometry, kinematics, and inclination of the BLR (Peterson, 2014).

Over the past decades, RM campaigns have been conducted for several tens of nearby AGN, primarily using optical broad lines such as H$\beta$ (e.g. Peterson, 1993, 2014). These measurements established the empirical BLR radius–luminosity ($R$–$L$) relation (e.g. Kaspi et al., 2000, 2005; Bentz et al., 2013), which allows RM-based black hole masses to be extrapolated to large AGN samples through single-epoch spectroscopy. As a result, single-epoch virial mass estimators are now widely used to infer SMBH masses for hundreds of thousands of AGN across cosmic time (e.g. Shen et al., 2011; Dalla Bontà et al., 2020), also enabling studies of the evolution of SMBH–galaxy scaling relations (e.g. Trujillo et al., 2004; Bennert et al., 2010). Beyond mass measurements, RM has also revealed the complex structure of the BLR.

Despite their success, virial mass estimates are subject to systematic uncertainties. In particular, the virial factor $f$ is not directly measurable for most AGN and has traditionally been calibrated by matching RM AGN to the $M_{\rm BH}$–$\sigma_\star$ relation of quiescent galaxies (Onken et al., 2004; Bennert et al., 2011; Park et al., 2012). An alternative promising approach is provided by BLR dynamical modelling, which fits RM data directly with physically motivated models and yields black hole masses independent of an assumed virial factor (e.g. the CARAMEL code, Brewer et al., 2011; Pancoast et al., 2011; Williams et al., 2018, 2020; Villafaña et al., 2022; Wang et al., 2026). Beyond mass measurements, RM has revealed the complex structure of the BLR: multi-line and velocity-resolved campaigns demonstrate clear radial stratification, with high-ionization lines responding on shorter timescales than low-ionization lines (e.g. Clavel et al., 1991; Grier et al., 2013; De Rosa et al., 2015), while dynamical modelling reveals diverse kinematic signatures including rotation, inflow, and outflow (e.g. Williams et al., 2020; Villafaña et al., 2022; Wang et al., 2026). Increasing evidence suggests that $f$ is not universal but depends on BLR structure, inclination, and accretion properties (e.g. Villafaña et al., 2023, 2026). The physical validity of BLR dynamical modelling is further supported by recent works like Bennert et al. (2026), who show that the inner AGN orientation as traced by CARAMEL-derived BLR inclination angles is uncorrelated with the large-scale host-galaxy disc, consistent with the BLR being physically linked to the inner accretion flow rather than to the host galaxy structure.

### 1.1. The Hubble constant tension and independent distance measurements

The Hubble constant $H_0$ sets the absolute distance scale of the Universe and quantifies the present-day cosmic expansion rate. Despite major advances in observational cosmology, measurements of $H_0$ obtained using early-Universe probes and late-Universe distance indicators remain in significant and persistent disagreement. This so-called "Hubble tension" has emerged as one of the most robust and long-standing discrepancies in modern cosmology, with a statistical significance currently at the 4–6$\sigma$ level depending on the datasets and assumptions adopted (Di Valentino et al., 2021).

Early-Universe determinations of $H_0$ are primarily based on observations of the cosmic microwave background (CMB), interpreted within the framework of the standard flat $\Lambda$CDM cosmological model. The *Planck* 2018 analysis gives $H_0 = 67.27 \pm 0.60$ km s$^{-1}$ Mpc$^{-1}$ for a six-parameter $\Lambda$CDM model (Planck Collaboration, 2020). Including CMB lensing information produces a nearly identical result, $H_0 = 67.36 \pm 0.54$ km s$^{-1}$ Mpc$^{-1}$. Independent CMB experiments broadly support this low value: analyses combining ACT and WMAP data find $H_0 = 67.6 \pm 1.1$ km s$^{-1}$ Mpc$^{-1}$, while SPT-3G reports $H_0 = 68.8 \pm 1.5$ km s$^{-1}$ Mpc$^{-1}$ under the same cosmological assumptions (Aiola et al., 2020; Dutcher et al., 2021).

In contrast, late-Universe measurements of $H_0$ rely on direct observations of distances and redshifts in the nearby Universe and are largely independent of early-Universe assumptions. The most precise such determination is provided by the SH0ES collaboration, which uses a Cepheid-calibrated Type Ia supernova (SNIa) distance ladder. The latest SH0ES result, obtained with 42 SNIa, yields $H_0 = 73.2 \pm 1.3$ km s$^{-1}$ Mpc$^{-1}$, corresponding to a $4.2\sigma$ tension with the *Planck* value (Riess et al., 2021). Alternative late-Universe distance indicators broadly support a high value of $H_0$. Water megamasers in the accretion discs of nearby AGN provide purely geometric distances and yield, in the latest papers, values at odds with Planck estimates $H_0 = 73.9 \pm 3.0$ km s$^{-1}$ Mpc$^{-1}$ (Pesce et al., 2020) for a sample of six megamasers. Strong-lensing time-delay cosmography provides another independent route, with recent analyses on eight multiply-imaged quasars yielding $H_0 = 71.6^{+3.9}_{-3.3}$ km s$^{-1}$ Mpc$^{-1}$ (Tdcosmo Collaboration et al., 2025).

The persistence of the Hubble tension across multiple independent methods suggests that it is unlikely to be explained by a single unidentified systematic affecting either early- or late-Universe measurements alone. If not attributable to observational systematics, the Hubble tension may point towards new physics beyond the standard $\Lambda$CDM model. A wide range of theoretical solutions have been proposed, including early dark energy, additional relativistic degrees of freedom, interacting dark sectors, and modifications of gravity, some of which can reduce the tension to the 1–2$\sigma$ level at the cost of introducing new parameters and degeneracies (Di Valentino et al., 2021). However, no single extension has yet emerged as a compelling and universally accepted resolution.

In this context, independent and purely geometric measurements of cosmological distances are of particular importance. Methods that do not rely on cross calibration can provide a critical cross-check of the distance-ladder determinations of $H_0$.

## 2. The SARM method: Combining reverberation mapping and spectroastrometry

The BLR emission of AGN provides a unique opportunity for geometric distance measurements. By combining reverberation mapping, which measures the physical size of the BLR, with interferometric measurements of its angular size, it is possible to infer angular-diameter distances in a calibration-independent manner. GRAVITY is a second-generation near-infrared beam combiner at the VLTI that coherently combines the light from all four 8-m Unit Telescopes, providing six simultaneous baselines and delivering micro-arcsecond astrometric precision (GRAVITY Collaboration et al., 2017, 2021). In its spectro-interferometric mode, GRAVITY measures wavelength-dependent differential phases across broad emission lines, which encode the photocenter displacement of the line-emitting gas relative to the continuum as a function of velocity. For a rotating BLR, this produces a characteristic S-shaped differential phase signal across the line profile, whose amplitude directly constrains the angular size of the BLR. By modelling both the line profile and the differential phase signal with a physically motivated BLR model, GRAVITY observations provide a robust measurement of the BLR angular radius, largely independent of assumptions about the source distance (e.g., Gravity Collaboration, 2018; Li et al., 2025; GRAVITY+ Collaboration et al., 2026). In the BLR model adopted by the GRAVITY/framework, the BLR is represented as an ensemble of clouds in the gravitational potential of the central black hole, distributed in a thick disc geometry with parameters including the mean and inner BLR radius, a radial shape parameter, disc opening angle, inclination, sky position angle, and BH mass (Gravity Collaboration, 2018). Early implementations assumed purely circular orbits and a globally linear BLR response, while more recent SARM analyses include radial-dependent responsivity (Li et al., 2025).

Given the linear size from RM and the angular size from GRAVITY interferometry measurements, the angular diameter distance can be written as

$$D_{\rm A} = \frac{R_{\rm BLR}}{\theta_{\rm BLR}}, \tag{2}$$

where $R_{\rm BLR}$ is the linear BLR size from RM and $\theta_{\rm BLR}$ is the angular BLR size from interferometric measurements. This combined approach, commonly referred to as the Spectroastrometry and Reverberation Mapping (SARM) method, was first demonstrated for the quasar 3C 273 by Wang et al. (2020) and then on NGC 3783 (GRAVITY Collaboration et al., 2021).

Li et al. (2025) have published an $H_0$ measurement combining four quasars with SARM and obtained $H_0 = 69^{+12}_{-10}$ km s$^{-1}$ Mpc$^{-1}$. They combined optical RM of the H$\beta$ line with near-infrared interferometric spectroastrometry of the Pa$\alpha$ or the Br$\gamma$ line (depending on the object redshift). This requires the assumption that the H$\beta$ and the Pa$\alpha$ (or Br$\gamma$) emitting regions trace the same underlying BLR structure. While both lines are hydrogen recombination lines and are expected to be broadly co-spatial, RM studies have established clear BLR stratification across emission lines (e.g. Bentz et al., 2010). Any mismatch between the regions probed by RM and spectroastrometry propagates directly into the distance estimate and represents a significant source of systematic uncertainty.

Using the same emission line for both RM and spectroastrometry would eliminate this source of systematic uncertainty and enable a purely geometric distance measurement. However, this requires near-infrared reverberation mapping of the same lines observed by GRAVITY, which is observationally challenging. Infrared hydrogen lines are intrinsically fainter than optical Balmer lines, with line ratios of about Pa$\alpha$/H$\beta \simeq 0.3$ and Br$\gamma$/H$\beta \simeq 0.03$ under typical conditions (Osterbrock and Ferland, 2006). As a result, Brackett lines in particular are difficult to monitor at the signal-to-noise and cadence required for reverberation mapping, especially for sources with modest variability amplitudes.

Despite these challenges, significant progress is now being made. Dedicated near-infrared RM campaigns targeting GRAVITY AGN have recently started using instruments such as Gemini/Flamingos-2 and IRTF/SpeX (Bennert et al. in prep, Signorini et al. in prep.), currently targeting nearby AGN (Ark 120, Mrk 509, NGC 3783, and Mrk 1239) selected as among the brightest GRAVITY targets with detectable near-infrared broad lines. For Ark 120, one of the brightest AGN on the sky (K $\simeq$ 9.8K), representative single-epoch SpeX spectra achieve an integrated SNR on the line flux of ~370 for Pa$\alpha$, ~180 for Pa$\beta$, and ~120 for Br$\gamma$ line. The Ark 120 campaign comprises 30 epochs at a cadence of ~4 days, while the ongoing campaigns on Mrk 509 and Mrk 1239 adopt longer cadences of 8–10 days, appropriate for their larger expected BLR sizes. These programmes provide a crucial proof of concept, and are demonstrating that the required SNR and cadence are achievable but only for a small number of exceptionally bright sources that already represent the limit of what current 4–8 m class facilities can routinely monitor. These campaigns also highlight why extending this approach to the broader GRAVITY+ sample remains out of reach with current instrumentation. Brackett lines are intrinsically weak and require high signal-to-noise ratios to detect variability at the few-percent level. Moreover, AGN variability is stochastic, and if a source exhibits only modest variability during the monitoring period, extracting a reliable lag from Br$\gamma$ becomes extremely challenging. Paschen lines, in particular Pa$\alpha$, are intrinsically stronger and therefore better suited for reverberation mapping, but they are accessible with GRAVITY only for AGN at moderate redshifts ($z \sim 0.1$–$0.3$). Flux calibration imposes an additional constraint. While optical RM usually relies on using narrow [O III] emission, near-infrared campaigns can use the narrow [S III]9531 line. This typically achieves ~3%–5% precision (Landt et al., 2019). This calibration floor limits the detectability of low-amplitude variability in NIR broad lines. Indeed, this calibration limit can be comparable to the variability signal itself (Zhang et al., 2019), limiting the ability to detect the flux variations needed to measure a reliable time lag.

### 2.1. GRAVITY+ and the emerging interferometric AGN sample

The ongoing GRAVITY+ upgrade aims to significantly extend the capabilities of the instrument by improving sensitivity and sky coverage through the implementation of laser guide star adaptive optics and off-axis fringe tracking (Gravity+ Collaboration et al., 2022). As per December 2025, the upgrades regarding the off-axis fringe-tracking mode, the new adaptive optics systems, and the laser guide stars are in commissioning phase (GRAVITY+ Collaboration et al., 2025a; GRAVITY+ Collaboration et al., 2025b).

GRAVITY+ is expected to be capable to observe several hundreds AGN with $K$-band magnitudes (Vega) up to $K \simeq 13$, spanning a wide range of redshifts and luminosities. These observations will provide measurements of BLR angular sizes and velocity-resolved photocentre shifts for emission lines such as Pa$\alpha$ and Br$\gamma$ for tens of objects up to redshift $z \sim 0.2$, enabling detailed studies of BLR geometry and black hole mass estimates based on spatially resolved kinematics. However, performing infrared RM on these objects to obtain an $H_0$ estimate would be very difficult with current facilities. A decrease of two magnitudes in K-band brightness corresponds to a factor of $\sim 6.3$ reduction in flux. If a K = 9 AGN requires a one-hour exposure on a 3 m-class telescope to achieve sufficient signal-to-noise on Br$\gamma$, an otherwise similar AGN with K = 11 would require approximately 6–7 h per epoch. Such exposure times are incompatible with the cadence and duration required for reverberation mapping campaigns, effectively restricting same-line SARM experiments to only the very brightest AGN with existing infrared instrumentation.

## 3. SHARP and NEXUS: Enabling near-infrared reverberation mapping

As discussed above, infrared RM is presently feasible for only a handful of very bright AGN, it remains challenging even for these sources, and it would not be feasible for objects observed in the future by GRAVITY+. SHARP is a proposed multi-mode near-infrared spectrograph for the Extremely Large Telescope, designed to fully exploit the collecting area of the 39 m aperture and the wide-field correction provided by the adaptive optics system. In particular, the NEXUS multi-object spectrograph mode would address the limitations of current infrared RM campaign. In this section we discuss how the scenario of AGN infrared RM would be changed by SHARP and the implications for the SARM method and the measurement of the $H_0$ constant.

### 3.1. Comparison with current facilities: the case of Ark 120

The feasibility of a SARM programme depends on the ability to obtain near-infrared spectra with sufficient signal-to-noise ratio (SNR), cadence, and spectrophotometric stability over extended monitoring periods. For reverberation mapping of broad infrared lines such as Br$\gamma$ and Pa$\alpha$, a SNR of at least~20–30 per resolution element is typically required.

Ark 120 is a prototypical nearby type 1 AGN and one of the brightest GRAVITY targets, with $K \simeq 9.8$ (Vega). Near-infrared reverberation mapping of Br$\gamma$ for such an object with existing facilities, such as IRTF/SpeX, typically requires exposure times of approximately one hour per epoch to reach SNR ~25 at moderate spectral resolution ($R \sim 2000$) under good atmospheric conditions (airmass $\lesssim 1.5$, stable precipitable water vapour). These observations are seeing-limited.

The collecting area of the ELT exceeds that of IRTF by a factor of ~150; in the photon-noise–limited regime, the S/N scales as S/N $\propto \sqrt{A\,t}$, implying that an exposure time of $t_{\rm SHARP} \sim 20$–$30$ s would deliver a S/N comparable to a one-hour IRTF exposure for a source like Ark 120. Furthermore, the adaptive optics system reduces the effective sky background per resolution element by more than an order of magnitude compared to seeing-limited observations. As a result, the effective gain for infrared RM with SHARP exceeds what is expected from aperture scaling alone. This dramatic gain enables a fundamentally different observing strategy. Short SHARP observations can be scheduled as filler programmes, allowing regular monitoring over months to years with minimal time investment. While each science exposure requires only tens of seconds for bright GRAVITY targets, total visit durations will be dominated by telescope preset and AO system acquisition overheads, estimated at ∼15–20 min per epoch based on experience with MCAO systems at existing 8-m class facilities. Such visits are well-suited to scheduling as filler observations within the ELT queue, allowing regular monitoring of bright AGN with modest total time investment over months to years.

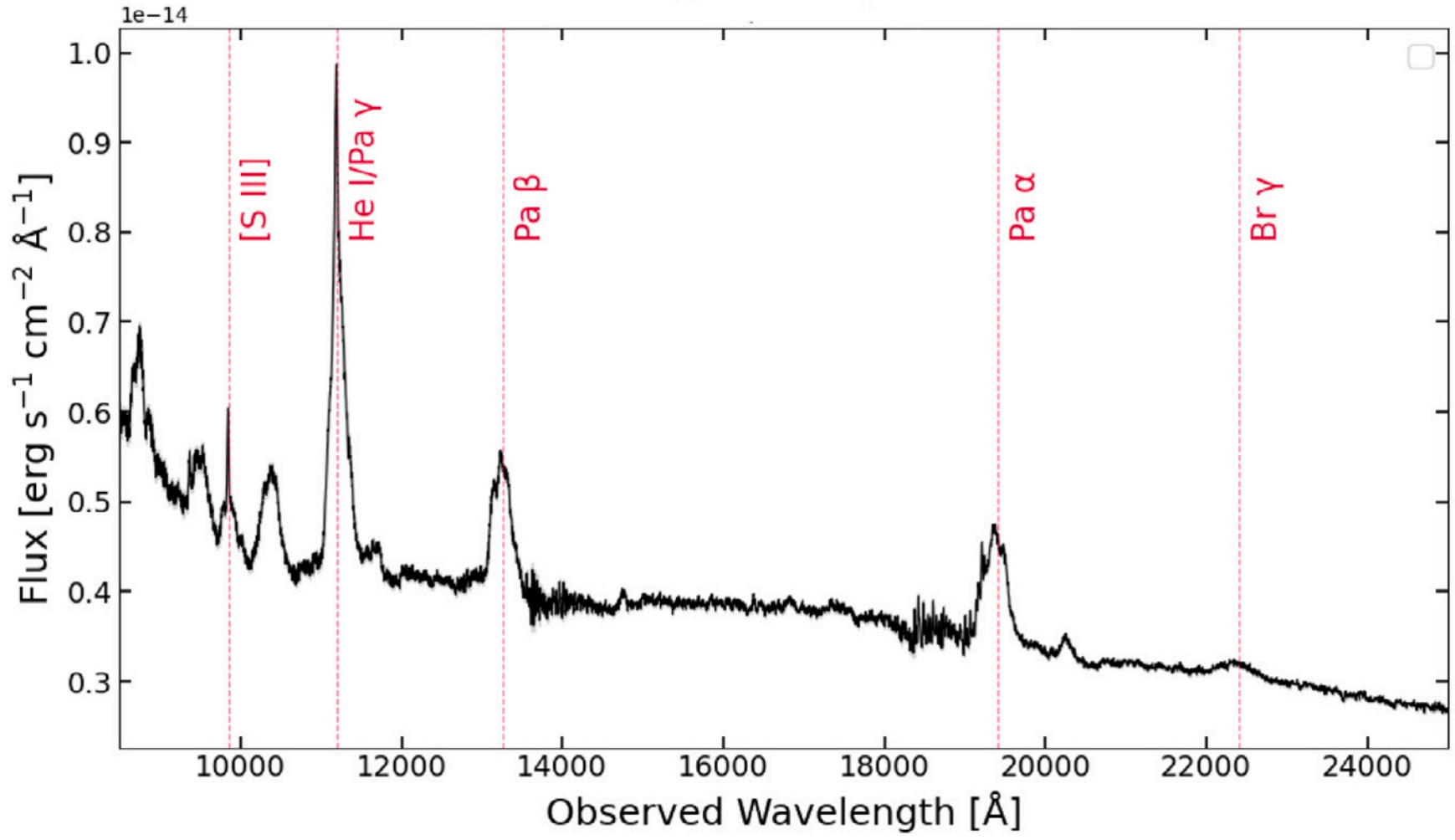

**Fig. 1.** Infrared spectrum of the AGN Ark 120 obtained at IRTF on 12/01/2024. In red, the strong Paschen lines and the [SIII] line that is used for flux calibration. A SNR or ∼25 on Br$\gamma$ has been obtained with about one hour exposure, for an object (Ark 120) with K magnitude (Vega) of ∼9.8. Fainter objects observed by GRAVITY+ would not be observable with this setup.

Given the stochastic nature of AGN variability, long baseline monitoring maximizes the probability of capturing active flux variation episodes. If a strong flux variation is detected, that could trigger a more high-cadence monitoring so that a reliable time lag for infrared lines could be measured, optimizing telescope time.

We used the SHARP Exposure Time Calculator (ETC v0.6) to estimate the feasibility of near-IR spectroscopic monitoring of a bright type 1 AGN representative of the current GRAVITY sample. We used Ark 120 and the spectral information we derived from the on-going IRTF monitoring (PI V. Bennert, see also Fig. 1), namely the integrated line flux and line width for the Br$\gamma$ $F_{\rm line} = 6.9 \times 10^{-14}\,{\rm erg\,s^{-1}\,cm^{-2}}$, FWHM = 442 Å. We adopted the NEXUS mode with $R = 2000$, MCAO correction, and a 200 mas slit. For a single 30 s exposure, the ETC predicts S/N $\simeq$ 270 per resolution element at the line wavelength, corresponding to the continuum level, with the observation being strongly source-dominated (target counts $\sim 7.5 \times 10^4\,e^-$ per exposure versus sky background $\sim 7.2 \times 10^2\,e^-$). Therefore for bright GRAVITY targets SHARP can deliver RM-quality spectroscopy in exposures of order tens of seconds, enabling high-cadence monitoring and efficient filler programmes.

To assess the feasibility of near-infrared reverberation mapping for fainter AGN that will be accessible with GRAVITY+, we extended the Ark 120 ETC simulation to a representative hypothetical object at redshift $z = 0.1$ with $K = 12$ (Vega). This choice is motivated by the expected magnitude range of the GRAVITY+ sample and allows a direct comparison with current facilities. Starting from the Ark 120 case, we rescaled the Br$\gamma$ emission-line properties under conservative and physically motivated assumptions. We assumed that the intrinsic broad-line velocity width remains unchanged, such that the observed-frame FWHM scales linearly with wavelength. Adopting the Ark 120 Br$\gamma$ FWHM of 442 Å at $\lambda_{\rm obs} \simeq 21661$ Å, this yields an observed-frame FWHM of ∼486 Å at $z = 0.1$, corresponding to a velocity width of $\sim 6000\,{\rm km\,s^{-1}}$. The line centre was shifted to $\lambda_{\rm obs} = 21661(1+z) \simeq 23827$ Å. The integrated Br$\gamma$ line flux was rescaled assuming a constant equivalent width and accounting for both the fainter near-infrared continuum and luminosity-distance dimming. Relative to Ark 120 ($K \simeq 9.8$, $z \simeq 0.03$), a source with $K = 12$ is fainter by $\Delta K \simeq 2.2$ mag, corresponding to a factor of ∼7.6 decrease in continuum flux. In addition, the increase in luminosity distance from ∼140 Mpc to ∼460 Mpc introduces a further $(D_{L,{\rm Ark}}/D_{L,z=0.1})^2 \simeq 0.09$ dimming. Combining these effects yields an expected integrated Br$\gamma$ flux of $F_{{\rm Br}\gamma} \simeq 8 \times 10^{-16}\,{\rm erg\,s^{-1}\,cm^{-2}}$, which we adopted as input to the ETC. We then used the SHARP ETC with the same conditions as described above. For a point-like source with the parameters above, the ETC predicts a signal-to-noise ratio S/N $\simeq$ 50 at the Br$\gamma$ wavelength for a total exposure time of 720 s, split into 24 exposures of 30 s each.

Finally, we considered the limiting case of AGN at the faint end of the GRAVITY+ capabilities, adopting a representative source with $K = 13$ at $z = 0.1$. Using the same assumptions as above for the Br$\gamma$ emission line (observed-frame FWHM of 486 Å and an integrated line flux $F_{{\rm Br}\gamma} = 3.2 \times 10^{-16}\,{\rm erg\,s^{-1}\,cm^{-2}}$), we used the SHARP ETC to determine the exposure time required to reach the minimum signal-to-noise ratio needed for reverberation mapping. For a total integration time of 1050 s (35 exposures of 30 s), the ETC predicts S/N $\simeq$ 25 at the Br$\gamma$ wavelength. In this regime, the observations are background dominated, with sky counts exceeding target counts per exposure, but the required exposure time remains modest. An integration of ∼15–20 min per epoch is therefore sufficient to achieve RM-quality spectroscopy even for AGN at the GRAVITY+ sensitivity limit. Beyond the signal-to-noise requirements, several physical effects add complexity to near-infrared broad-line measurements. The broad-line profile is rarely a simple symmetric shape: outflowing or inflowing gas components can produce line asymmetries and non-reverberating flux contributions that dilute the variability signal and potentially bias the measured time lag (e.g. Bentz et al., 2010; Williams et al., 2020). Profile variability between epochs can further complicate the extraction of a reliable integrated line light curve. In the near-infrared, the continuum is also more complex than in the optical: in addition to the AGN power-law component, contributions from host galaxy starlight and hot dust emission must be carefully modelled and subtracted before the broad-line flux can be measured. Finally, broad Paschen and Brackett lines can be partially blended with narrow emission features or affected by imperfect sky subtraction residuals, particularly in the H and K bands where telluric absorption is significant. While these effects are present in optical RM campaigns as well, they are generally more challenging to handle in the near-infrared due to the lower line-to-continuum ratios and the harsher atmospheric background. Spectral decomposition techniques and the use of simultaneous multi-object observations to track instrumental and atmospheric systematics will be important to mitigate these complications.

### 3.2. Spectrophotometric precision and multi-object calibration

Achieving precise time lags requires not only high S/N but also excellent relative flux calibration between epochs. The use of prominent infrared narrow lines (e.g. [S III] $\lambda$9531) can deliver calibration precisions of ∼3%–5% (Landt et al., 2019), but this level of accuracy is often insufficient to robustly detect low-amplitude variability in infrared broad lines. SHARP's NEXUS multi-object capability enables the simultaneous observation of the AGN and multiple stars in the same field of view, providing a stable reference against variations in seeing, transparency, and slit losses. This approach has been demonstrated successfully in optical reverberation mapping of quadruply lensed quasars, where differential calibration against field stars yields sub-percent flux precision (Williams et al., 2021), as can be seen in Fig. 2. Applying the same strategy systematically in the near-infrared with SHARP is expected to deliver relative flux accuracies of $\lesssim 1\%$, substantially improving the precision and robustness of Br$\gamma$ and Pa$\alpha$ lag measurements compared to current infrared RM efforts. The importance of reaching this level of precision becomes clear when considering that the 3C 273 campaign of Zhang et al. (2019) measured a broad-line flux variability amplitude of ∼5%, comparable to the 3%–5% calibration floor of current near-infrared instruments. With calibration uncertainties at or above the variability signal, robust lag measurements become extremely difficult. Looking forward, simulations by Songsheng et al. (2021) show that GRAVITY+ can in principle constrain $H_0$ to ∼2% by observing ∼60 targets, provided per-source RM uncertainties remain at the ∼10% level. The $\lesssim$1% flux precision enabled by SHARP's simultaneous multi-object mode would place the calibration floor well below the variability signal, ensuring that the RM component does not become the dominant source of uncertainty as GRAVITY+ delivers a substantially larger and fainter AGN sample.

### 3.3. Implications for SARM

The capabilities of SHARP, particularly in its NEXUS multi-object mode, would transform the potential of the SARM method. By combining the unprecedented collecting area of the ELT and adaptive optics spectroscopy, SHARP can address the two major limitations of current infrared RM efforts: sensitivity and calibration. The dramatic gain in photon-collecting power and background suppression means that exposures of only tens of seconds on SHARP can achieve the same signal-to-noise ratio as hour-long integrations on 3–4 m class telescopes. This efficiency enables a fundamentally different observing strategy: low-cost, long-term monitoring of bright AGN already observed by GRAVITY. Such monitoring can be scheduled as filler observations, ensuring continuous coverage over months to years. When significant variability is detected, high-cadence follow-up can be triggered to secure robust lag measurements for infrared lines such as Br$\gamma$ and Pa$\alpha$. This approach will give us precise time lags for a substantial sample of nearby AGN without the prohibitive time costs that currently limit infrared RM. Furthermore, GRAVITY+ will expand interferometric AGN studies to fainter sources (up to K ∼13), which are inaccessible to current infrared RM facilities. SHARP overcomes this, making near-infrared RM feasible within practical exposure times. This capability opens the door to systematic RM campaigns for tens of AGN across a wide luminosity and redshift range, fully exploiting the GRAVITY+ sample. The GRAVITY+ sample accessible to near-infrared RM with SHARP spans $K \simeq 9$–$13$ (Vega) and redshifts $z \lesssim 0.2$, corresponding to AGN luminosities of roughly $\lambda L_\lambda(5100A) \sim 10^{43} - 10^{45}$erg/s. Via the BLR radius–luminosity relation (Bentz et al., 2013), this luminosity range implies characteristic BLR sizes of order ∼10–200 light-days and expected Br$\gamma$ or Pa$\alpha$ time lags of weeks to months in the observed frame. This sets the natural cadence and campaign duration requirements: monitoring at ∼5–15 day cadence over 1–3 years would be required to robustly sample the BLR response across the full sample. For the brightest targets (K $\lesssim$ 11), such campaigns are feasible as low cost filler programmes with SHARP as described in Section 3.1. For fainter sources approaching the GRAVITY+ sensitivity limit (K ∼13), the modest per epoch exposure times of ∼15–20 min estimated in Section∼3.1 make systematic monitoring still practical. This opens the door to a sample of several tens of AGN for which same line SARM distances and therefore an independent measurement of $H_0$ can be obtained.

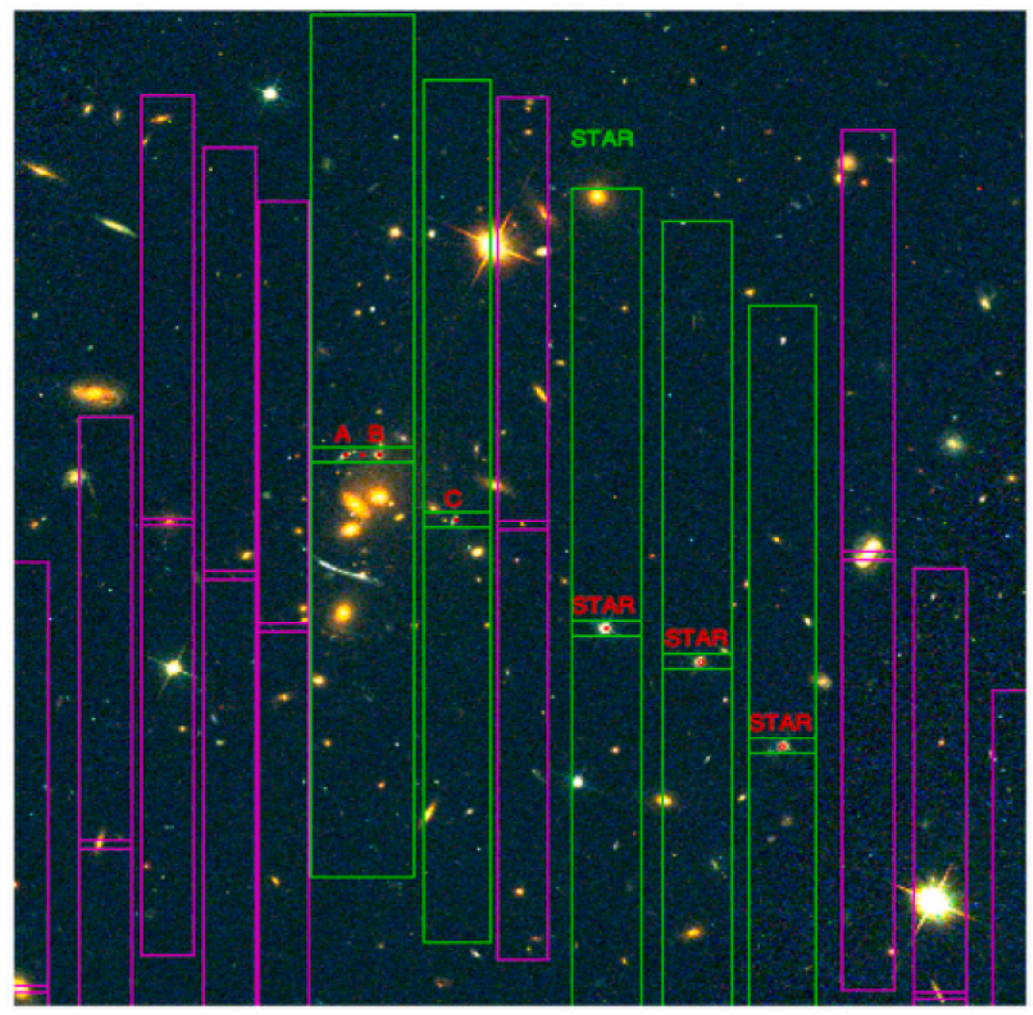

**Fig. 2.** From Williams et al. (2021), an example of a multi-object instrument (in this case GMOS-MOS) used to provide cross-observation flux calibration for reverberation mapping. The use of field stars as flux reference allowed a flux precision $\leq 1\%$.

These advances have two main implications:

(1) Measurement of the Hubble constant: With tens of AGN providing precise infrared lags, combined with GRAVITY+ spectroastrometry, SARM can deliver distances in a completely geometric, calibration-independent manner. This will enable an independent determination of $H_0$, providing a critical cross-check on the current Hubble tension.

(2) Understanding of BLR physics and SMBH masses: Simultaneous RM and interferometric observations in the same infrared lines will allow detailed mapping of BLR structure and kinematics. This synergy will refine black hole mass estimates, helping us understand more about the virial factor possible dependence on other physical quantities. It will also allow us to test BLR stratification in the infrared, and track structural evolution over time, offering new insights into accretion physics and AGN evolution.

It is worth summarizing the expected final accuracy on $H_0$ from the combined SHARP and GRAVITY+ programme, considering all major sources of uncertainty. The current state of the art has been obtained for four quasars analysed with SARM by Li et al. (2025). They obtain an estimate of $H_0 = 69 \pm 11\,\mathrm{km\,s^{-1}\,Mpc^{-1}}$, corresponding to a ∼16% precision. The uncertainty is dominated by the limited sample size and the differential phase uncertainty of current GRAVITY observations. For individual sources, the $D_A$ uncertainty budget includes contributions from: (i) the interferometric angular size measurement ($\sigma_\theta/\theta$), currently ∼15%–20% per source and expected to improve to ≲10% with GRAVITY+ thanks to better phase sensitivity and access to stronger lines; (ii) the RM-based physical size measurement ($\sigma_R/R$), which depends on the time lag precision and the flux calibration stability; and (iii) modelling systematics associated with the BLR geometry assumptions. Since $\sigma_{H_0}/H_0 \propto 1/\sqrt{N}$ for a sample of $N$ statistically independent measurements, the precision on $H_0$ scales as the per-source uncertainty divided by $\sqrt{N}$. Simulations by Songsheng et al. (2021) show that observing ∼60 targets with per-source RM uncertainties at the ∼10% level would constrain $H_0$ to ∼2%, a level competitive with current independent probes and sufficient to meaningfully contribute to resolving the Hubble tension. The GRAVITY+ sample of tens of AGN with $K \lesssim 13$ and $z \lesssim 0.2$, combined with SHARP near-infrared RM campaigns delivering ≲1% flux calibration precision, would bring this goal within reach. Residual systematic uncertainties from BLR modelling assumptions, line stratification, and the stochastic nature of AGN variability will ultimately limit the accuracy of individual distance measurements and must be characterized through the growing sample itself.

In short, SHARP would transform infrared RM from a niche capability into a systematic tool for cosmology and AGN science.

## CRediT authorship contribution statement

**Matilde Signorini:** Writing – review & editing, Writing – original draft, Formal analysis, Data curation, Conceptualization. **Vardha N. Bennert:** Writing – review & editing, Methodology, Investigation, Data curation, Conceptualization. **Elena Dalla Bontà:** Writing – review & editing, Conceptualization. **Federica Ricci:** Writing – review & editing, Validation, Conceptualization. **Lizvette Villafaña:** Writing – review & editing, Methodology, Investigation, Conceptualization.

## Declaration of competing interest

The authors declare that they have no known competing financial interests or personal relationships that could have appeared to influence the work reported in this paper.

## Acknowledgements

MS acknowledges support through the European Space Agency (ESA) Research Fellowship Programme.

## Data availability

Data will be made available on request.